\documentclass[twocolumn,prb,showpacs,superscriptaddress,preprintnumbers,amsmath,amssymb]{revtex4}

\usepackage[dvips]{graphicx}
\usepackage{dcolumn}
\usepackage{bm}
\begin{document}
\title{Magnetic versus crystal field linear dichroism in NiO thin films}

\author{M. W. Haverkort}
 \affiliation{ II. Physikalisches Institut, Universit\"{a}t zu K\"{o}ln, Z\"{u}lplicher Str. 77, 50937 K\"{o}ln, Germany}
\author{S. I. Csiszar}
 \affiliation{ MSC, University of Groningen, Nijenborgh 4, 9747 AG Groningen, The Netherlands}
\author{Z. Hu}
 \affiliation{ II. Physikalisches Institut, Universit\"{a}t zu K\"{o}ln, Z\"{u}lplicher Str. 77, 50937 K\"{o}ln, Germany}
\author{S. Altieri}
 \affiliation{ INFM - National Center on Nanostructures and Biosystems at Surfaces (S$^{3}$), I-41100 Modena, Italy}
\author{A. Tanaka}
 \affiliation{ Department of Quantum Matter, ADSM, Hiroshima University, Higashi-Hiroshima 739-8530, Japan}
\author{H. H. Hsieh}
 \affiliation{National Synchrotron Radiation Research Center, 101 Hsin-Ann Road, Hsinchu 30077, Taiwan}
\author{H.-J. Lin}
 \affiliation{National Synchrotron Radiation Research Center, 101 Hsin-Ann Road, Hsinchu 30077, Taiwan}
\author{C. T. Chen}
 \affiliation{National Synchrotron Radiation Research Center, 101 Hsin-Ann Road, Hsinchu 30077, Taiwan}
\author{T. Hibma}
 \affiliation{ MSC, University of Groningen, Nijenborgh 4, 9747 AG Groningen, The Netherlands}
\author{L. H. Tjeng}
 \affiliation{ II. Physikalisches Institut, Universit\"{a}t zu K\"{o}ln, Z\"{u}lplicher Str. 77, 50937 K\"{o}ln, Germany}

\date{\today}

\begin{abstract}
We have detected strong dichroism in the Ni $L_{2,3}$ x-ray absorption spectra
of monolayer NiO films. The dichroic signal appears to be very similar to the
magnetic linear dichroism observed for thicker antiferromagnetic NiO films. A
detailed experimental and theoretical analysis reveals, however, that the
dichroism is caused by crystal field effects in the monolayer films, which is a
non trivial effect because the high spin Ni $3d^{8}$ ground state is not split
by low symmetry crystal fields. We present a practical experimental method for
identifying the independent magnetic and crystal field contributions to the
linear dichroic signal in spectra of NiO films with arbitrary thicknesses and
lattice strains. Our findings are also directly relevant for high spin $3d^{5}$
and $3d^{3}$ systems such as LaFeO$_{3}$, Fe$_{2}$O$_{3}$, VO, LaCrO$_{3}$,
Cr$_{2}$O$_{3}$, and Mn$^{4+}$ manganate thin films.
\end{abstract}


\maketitle

Magnetic linear dichroism (MLD) in soft-x-ray absorption spectroscopy (XAS) has
recently developed into one of the most powerful tools to study the magnetic
properties of antiferromagnetic thin films
\cite{Thole85,Sinkovic90,Kuiper93,Alders95,Alders98}. The contrast that one can
obtain as a result of differences in the magnitude and orientation of local
moments is essential to determine the spin anisotropy and important parameters
like the N\'{e}el temperature $(T_{N})$, as well as to map out spatially the
different magnetic domains that are present in antiferromagnetic films
\cite{Stohr98,Spanke98,Stohr99,Scholl00,Nolting00,Ohldag01a,Ohldag01b,Zhu01,Hille01}.
Such information is extremely valuable for the research and application of
magnetic devices that make use of exchange bias.

  \begin{figure}
    \includegraphics[width=0.45\textwidth]{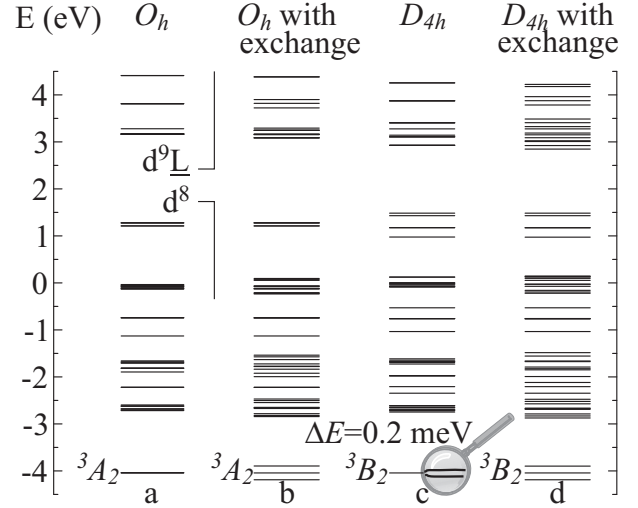}
    \caption{Energy level diagram for Ni$^{2+}$ ($3d^{8}$) in:
            (a) $O_{h}$ symmetry with $pd\sigma$=-1.29 and $10Dq$=0.85 eV;
            (b) $O_{h}$ symmetry with additional exchange field of 0.16 eV;
            (c) $D_{4h}$ symmetry with $pd\sigma$=-1.29, $10Dq$=0.85, and $Ds$=0.12 eV;
            (d) $D_{4h}$ symmetry with additional exchange.
            The $3d$ spin-orbit interaction is included, but the states are labelled
            as if the spin-orbit interaction was not present}\label{fig1}
  \end{figure}

Much of the modern MLD work has been focussed on NiO and LaFeO$_{3}$ thin
films, and the observed dichroism has been attributed entirely to magnetic
effects
\cite{Stohr98,Spanke98,Stohr99,Scholl00,Nolting00,Ohldag01a,Ohldag01b,Zhu01,Hille01}.
Other sources that could contribute to linear dichroism, however, such as
crystal fields of lower than octahedral symmetry, have been neglected or not
considered. Indeed, one would expect that such low symmetry crystal fields are
negligible for bulk-like NiO and LaFeO$_{3}$ films, and, more fundamentally,
that such crystal fields will not split the high-spin Ni $3d^{8}$ or Fe
$3d^{5}$ ground state. We have illustrated this insensitivity in Fig. 1 for the
Ni$^{2+}$ case, where the energy level diagram in an $O_{h}$ environment is
compared to that in a $D_{4h}$ point group symmetry \cite{SOremark}. In
contrast, an exchange field will split the Ni$^{2+}$ ground state into three
levels with $M_{S}$=-1,0,1 with an energy separation given by the exchange
coupling $J$, see Fig. 1. The basis for obtaining strong dichroism in the Ni
$L_{2,3}$ ($2p$$\rightarrow$$3d$) absorption spectra is that dipole selection
rules dictate which of the quite different final states can be reach and with
what probability for each of the initial states. The isotropic spectrum of each
of these three states will be the same, but each state with a different
$|M_{S}|$ value will have a different polarization dependence
\cite{Thole85,Sinkovic90,Kuiper93,Alders95,Alders98}. A completely analogous
argumentation can be given for the orbitally highly symmetric high spin
$3d^{5}$ and $3d^{3}$ cases, e.g. Mn$^{2+}$, Fe$^{3+}$, V$^{2+}$, Cr$^{3+}$,
Mn$^{4+}$.

In this paper we report on XAS measuments on single monolayer (ML) NiO films
which are grown on a Ag(100) substrate and capped by a 10 ML MgO(100) film. We
have observed strong linear dichroism in the Ni $L_{2,3}$ spectra, very similar
to that measured for thicker NiO films. From a detailed theoretical and
experimental analysis, however, we discovered that the dichroism can not be
attributed to the presence of some form of magnetic order, but entirely to
crystal field effects. The analysis provides us also with a practical guide of
how to disentangle quantitatively the individual contributions to the linear
dichroic signal, i.e. the contribution from magnetic interactions versus that
from low symmetry crystal fields. This is important for a reliable
determination of, for instance, the spin moment orientation in NiO as well as
LaFeO$_{3}$, Fe$_{2}$O$_{3}$, VO, LaCrO$_{3}$, Cr$_{2}$O$_{3}$, and Mn$^{4+}$
manganate ultra thin films, surfaces and strained films, where the low symmetry
crystal field splittings may not be negligible as compared to the exchange
field energies.

The polarization dependent XAS measurements were performed at the Dragon
beamline of the National Synchrotron Radiation Research Center in Taiwan. The
spectra were recorded using the total electron yield method in an XAS chamber
with a base pressure of 3x10$^{-10}$ mbar. The photon energy resolution at the
Ni $L_{2,3}$ edges ($h\nu \approx 850-880$ eV) was set at 0.3 eV, and the
degree of linear polarization was $\approx 98 \%$. A NiO single crystal is
measured \textit{simultaneously} in a separate chamber upstream of the XAS
chamber in order to obtain a relative energy reference with an accuracy of
better than 0.02 eV. The 1 ML NiO film on Ag(100) was prepared in Groningen, by
using NO$_{2}$ assisted molecular beam epitaxy. Immediately after the NiO
growth, the sample was capped \textit{in-situ} with an epitaxial 10 ML MgO(100)
film. Reflection high energy electron diffraction (RHEED) intensity
oscillations recorded during growth of thicker films demonstrated the
layer-by-layer growth mode and provided an accurate thickness calibration
\cite{Altierithesis,Altieri99}.

  \begin{figure}
    \includegraphics[width=0.45\textwidth]{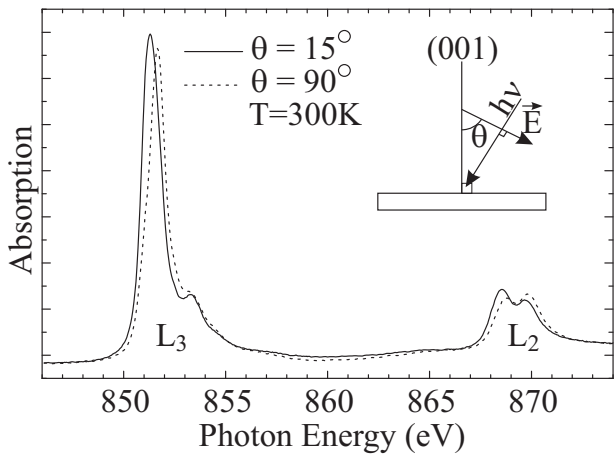}
    \caption{Experimental polarization dependent Ni $L_{2,3}$ XAS of 1 ML
             NiO(100) on Ag(100) covered with MgO(100). $\theta$ is the angle
             between the light polarization vector and the (001) surface normal
             ($\theta$=$90^{\circ}$ means normal light incidence).
    }\label{fig2}
  \end{figure}

Fig. 2 shows the polarization dependent Ni $L_{2,3}$ XAS spectra of the 1 ML
NiO film, taken at room temperature. The angle between the light polarization
vector and the (001) surface normal is given by $\theta$ ($\theta = 90^{\circ}$
means normal light incidence). The general lineshape of the spectra is very
similar to that of thicker NiO films and bulk NiO \cite{Alders98}.

  \begin{figure}
    \includegraphics[width=0.45\textwidth]{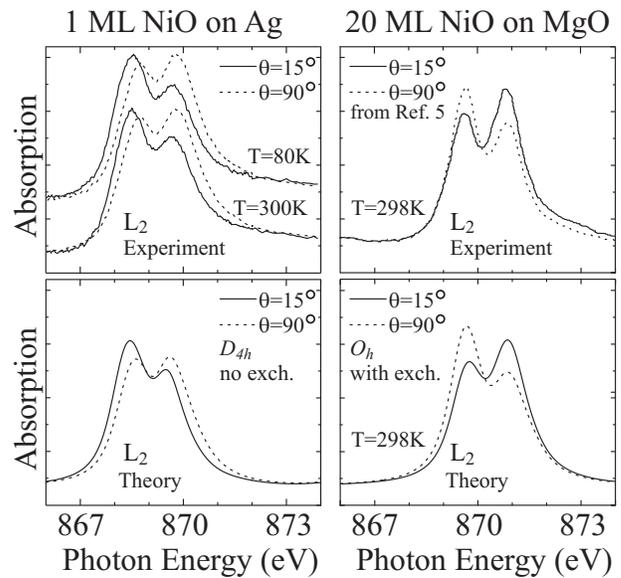}
    \caption{Polarization dependence of the Ni $L_{2}$ XAS of 1 ML NiO on
    Ag(100) covered by MgO(100). The 20 ML NiO on MgO spectra are taken from Ref. 5.
    The theoretical spectra for the 1 ML NiO
    are calculated in $D_{4h}$ symmetry without
    exchange, and for the 20 ML in $O_{h}$ with exchange.}
  \end{figure}

Fig. 3 presents a close-up of the $L_{2}$ edge, the region most often used to
measure the magnitude of the magnetic linear dichroic effect in
antiferromagnetic NiO films
\cite{Stohr98,Spanke98,Stohr99,Ohldag01a,Ohldag01b,Zhu01,Hille01}. The spectra
of the 1 ML NiO film show a very clear polarization dependence. This linear
dichroic effect is as strong as that for a 20 ML NiO film grown on MgO(100)
(taken from Alders \textit{et al.} \cite{Alders98}), albeit with an opposite
sign, as can be seen from Fig. 3. Relying on the analysis by Alders \textit{et
al.} \cite{Alders98} for the antiferromagnetic 20 ML film, one may be tempted
to conclude directly that the spin orientation in the 1 ML film is quite
different to that of the 20 ML film, i.e. that the spins for the 1 ML would be
lying more parallel to the interface while those of the thicker films are
pointing more along the interface normal. However, Alders \textit{et al.}
\cite{Alders98} have also shown that the magnetic ordering temperature of NiO
films decreases strongly if the film is made thinner. In fact, for a 5 ML NiO
film on MgO(100), it was found that $T_{N}$ is around or below room
temperature, i.e. that no linear dichroism can be observed at room temperature.
A simple extrapolation will therefore suggest that 1 ML NiO will not be
magnetically ordered at room temperature. This is in fact supported by the 80 K
data of the 1 ML NiO on Ag(100) as shown in Figs. 2 and 3: the spectra and the
dichroism therein are identical to those at 300 K, indicating that $T_{N}$ must
be at least lower than 80 K.

  \begin{figure}
    \includegraphics[width=0.45\textwidth]{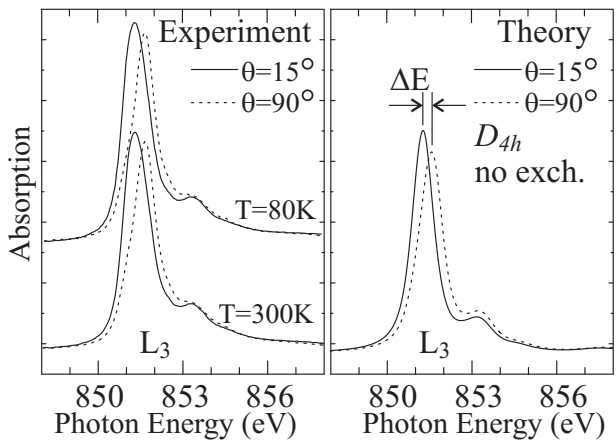}
    \caption{Theoretical and experimental polarization dependence of the Ni
    $L_{3}$-XAS of a 1 ML NiO(100) on Ag(100) covered with MgO(100). The
    theoretical spectra are calculated in $D_{4h}$ symmetry without
    exchange.}\label{fig4}
  \end{figure}

In order to resolve the origin of the linear dichroism in the 1 ML NiO system,
we now resort to the Ni $L_{3}$ part of the spectrum. A close-up of this region
is given in Fig. 4. We can easily observe that the strong polarization
dependence of the spectra is accompanied by an energy shift $\Delta E$ of 0.35
eV in the main peak of the $L_{3}$ white line. This shift seems small compared
to the 852 eV photon energy being used, but it is very reproducible and well
detectable since the photon energy calibration is done with an accuracy of
better than 0.02 eV thanks to the simultaneous measurement of a NiO single
crystal reference. We now take this energy shift as an indicator for the
presence and strength of local crystal fields with a symmetry lower than
$O_{h}$, i.e. crystal fields that do not split the ground state but do alter
the energies of the XAS final states, and, via second order processes, also
causes spectral weight to be transferred between the various peaks as we will
show below.

To understand the Ni $L_{2,3}$ spectra quantitatively, we perform calculations
for the atomic $2p^{6}3d^{8} \rightarrow 2p^{5}3d^{9}$ transitions using the
same method as described earlier by Alders \textit{et al.} \cite{Alders98}, but
now in a $D_{4h}$ point group symmetry. The method uses the full atomic
multiplet theory and includes the effects of the solid. It accounts for the
intra-atomic $3d$-$3d$ and $2p$-$3d$ Coulomb and exchange interactions, the
atomic $2p$ and $3d$ spin-orbit couplings, the O $2p$ - Ni $3d$ hybridization
with $pd\sigma$ = -1.29 eV, and an $O_{h}$ crystal field splitting of $10Dq$ =
0.85 eV. The local symmetry for the Ni ion sandwiched between the Ag(100)
substrate and the MgO(100) film is in principle $C_{4v}$, but for $d$ electrons
one can ignore the odd part of the crystal field, so that effectively one can
use the tetragonal $D_{4h}$ point group symmetry. As we will explain below, the
$D_{4h}$ parameters $Ds$ and $Dt$ \cite{Ballhausen62} are set to 0.12 and 0.00
eV, respectively, and the exchange field (the molecular field acting on the
spins) to zero. The calculations have been carried out using the XTLS 8.0
programm\cite{Tanaka94}.

The right panel of Fig. 4 shows the calculated $L_{3}$ spectrum for the light
polarization vector perpendicular and parallel to the $C_{4}$ axis ($\theta =
90^{\circ}$ and $\theta = 0^{\circ}$, respectively). One can clearly see that
the major experimental features are well reproduced, including the 0.35 eV
energy shift between the two polarizations. This shift can be understood in a
single electron picture. The ground state has the
$\underline{3d}_{x^{2}-y^{2}}\underline{3d}_{z^{2}}$ configuration, where the
underline denotes a hole. The final state has a
$2p^{5}\underline{3d}_{x^{2}-y^{2}}$ or $2p^{5}\underline{3d}_{z^{2}}$
configuration. For $z$ polarized light the $3d_{z^{2}}$ state can be reached,
but the $3d_{x^{2}-y^{2}}$ can not, and the final state will be of the form
$2p^{5}\underline{3d}_{x^{2}-y^{2}}$. For $x$ polarized light the final state
will be of the form $2p^{5}\underline{3d}_{z^{2}-y^{2}}$$=
$$\sqrt{3/4}(2p^{5}\underline{3d}_{z^{2}})$$+
$$\sqrt{1/4}(2p^{5}\underline{3d}_{x^{2}-y^{2}})$. In a pure ionic picture, the
$2p^{5}\underline{3d}_{x^{2}-y^{2}}$ state will be 4$Ds$+5$Dt$ lower in energy
than the $2p^{5}\underline{3d}_{z^{2}}$ state. In the presence of the O $2p$ -
Ni $3d$ hybridization, we find that $Ds$=0.12 and $Dt$=0.00 eV reproduce the
observed 0.35 eV shift.

Going back to the $L_{2}$ edge, we can see in Fig. 3 that the calculations can
also reproduce very well the observed linear dichroism in the 1 ML NiO spectra.
In fact, one now could also see the same 0.35 eV shift at this edge, although
it is not as clear as in the $L_{3}$ edge. We would like to stress here that
the good agreement has been achieved without the inclusion of an exchange
splitting, i.e. the dichroism is solely due to the low symmetry crystal field
splitting. It is a final state effect and the change in the ratio between the
two peaks of the $L_{2}$ edge as a function of polarization can be understood
as follows. In $O_{h}$ symmetry the first peak is due to two final states, one
of $T_{2}'$ and one of $E_{1}'$ symmetry. The second peak is due to a final
state of $T_{1}'$ symmetry. All three states have the
$2p^{5}\underline{3d}_{{e}_{g}}$ configuration. If one reduces the crystal
field to $D_{4h}$ symmetry the peaks will split. The $T_{2}'$ state will split
into two states of $B_{2}'$ and $E_{1}'$ symmetry, the $E_{1}'$ into $A_{1}'$
and $B_{1}'$, and the $T_{1}'$ into $A_{2}'$ and $E_{1}'$. The energy splitting
can be measured, but this is much easier done using the $L_{3}$ edge. We note
that each of the two peaks in the L$_{2}$ edge will have a state of $E_{1}'$
symmetry, so that these two will mix and transfer spectral weight. This can be
seen with isotropic light, but will show up more pronounced as a linear
dichroic effect if polarized light is used.

In contrast to the 1 ML NiO case, the good agreement between theory and
experiment for the polarization dependent spectra of a 20 ML NiO film
\cite{Alders98} have been achieved by assuming the presence of an
antiferromagnetic order with an exchange field of about 0.16 eV in a pure local
$O_h$ symmetry. It is surprising and also disturbing that a low symmetry
crystal field could induce a spectral weight transfer between the two peaks of
the $L_{2}$ white line such that the resulting linear dichroism appears to be
very similar as a dichroism of magnetic origin. It is obvious that the ratio
between the two peaks can not be taken as a direct measure of the spin
orientation or magnitude of the exchange field in NiO films \cite{Alders98} if
one has not first established what the crystal field contribution could be.

  \begin{figure}
    \includegraphics[width=0.45\textwidth]{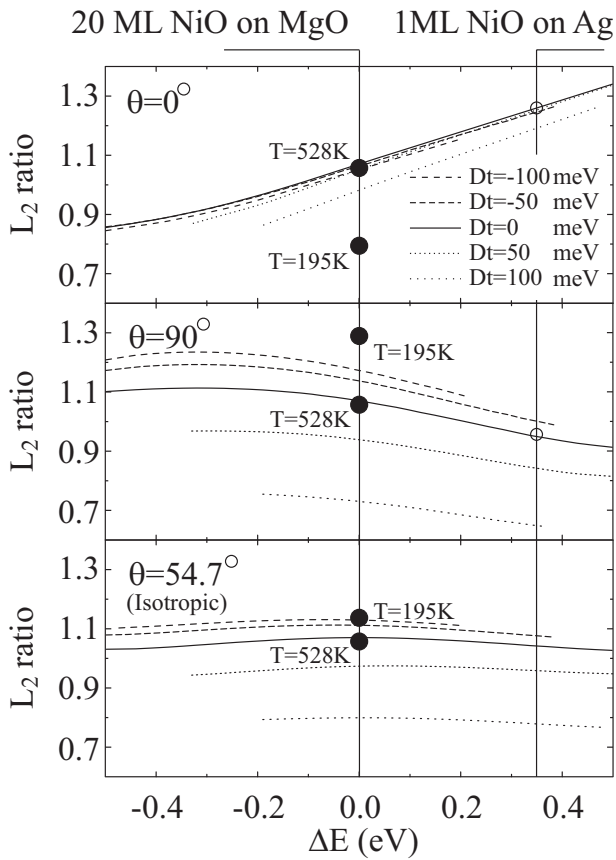}
    \caption{Calculated ratio of the two peaks in the $L_{2}$ edge as a
     function of $\Delta E$, which is the shift in energy of the $L_{3}$
     main peak in going from normal ($\theta = 90^{\circ}$) to grazing
     ($\theta = 0^{\circ}$) incidence of the linearly polarized light.
     The $L_{2}$ ratio is calculated for $\theta = 90^{\circ}$,
     $\theta = 0^{\circ}$, and for the isotropic spectrum.
     }\label{fig5}
  \end{figure}

We now can identify two strategies for finding out which part of the linear
dichroism is due to low symmetry crystal field effects. The first one is to
study the temperature dependence as we have done above. Here we have made use
of the fact that those crystal fields do not split the high spin ground state,
so that there are no additional states to be occupied with different
temperatures other than those already created by the presence of exchange
fields. Thus there should not be any temperature dependence in the crystal
field dichroism. The linear dichroism due to magnetism, however, is temperature
dependent and scales with $<$$M^2$$>$. By going to temperatures high enough
such that there is no longer any temperature dependence in the linear
dichroism, i.e. when all magnetic ordering has been destroyed, one will find
the pure crystal field induced dichroism.

The second strategy to determine the low symmetry crystal field contribution is
to measure carefully the energy shift $\Delta E$ in the main peak of the Ni
$L_{3}$ white line for $\theta$=$0^{\circ}$ vs. $\theta$=$90^{\circ}$. We now
calculate the ratio between the two peaks of the Ni $L_{2}$ edge as a function
of $\Delta E$, and the results are plotted in Fig. 5 for $\theta =90^{\circ}$,
$\theta =0^{\circ}$, and $\theta =54.7^{\circ}$ (isotropic spectrum). Since
$\Delta E$ is a function of $Ds$ and $Dt$ combined, we have carried out the
calculations with $Ds$ as a running variable for several fixed values of $Dt$,
and plotted the resulting $L_{2}$ ratios vs. $\Delta E$. We now can use Fig. 5
as a road map to determine how much of the linear dichroism in the Ni $L_{2}$
edge is due to crystal field effects and how much due to magnetism. We can see
directly that the 1 ML data lie on curves with the same $Dt$, meaning that the
measured $L_{2}$ ratios are entirely due to crystal fields. The same can also
be said for the 20 ML NiO on MgO at 528 K, which is not surprising since this
temperature is above $T_{N}$. However, for the 20 ML NiO at 195 K, one can see
that the data points do not lie on one of the $Dt$ curves (one may look for
larger $Dt$ curves, but this results in lineshapes very different from
experiment) or let alone on curves with the same $Dt$, indicating that one need
magnetism to explain the $L_{2}$ ratios. In other words, knowing the $L_{2}$
ratio and $\Delta E$ together allows us to determine the magnitude of the
exchange interaction and the orientation of the spin moments. It is best to use
the $\theta = 0^{\circ}$ spectra since here the $L_{2}$ ratio is determined
almost by $\Delta E$ alone and is not too sensitive to the individual values of
$Ds$ and $Dt$.

To conclude, we have observed strong linear dichroism in the 1 ML NiO on Ag,
very similar to the well known magnetic linear dichroism found for bulk like
antiferromagnetic NiO films.  The dichroism in the 1 ML, however, can not be
attributed to the presence of some form of magnetic order, but entirely to
crystal field effects. We provide a detailed analysis and a practical guide of
how to disentangle quantitatively the magnetic from the crystal field
contributions to the dichroic signal. This is important for a reliable
determination of, for instance, the spin moment orientation in NiO as well as
LaFeO$_{3}$, Fe$_{2}$O$_{3}$, VO, LaCrO$_{3}$, Cr$_{2}$O$_{3}$, and Mn$^{4+}$
manganate ultra thin films, surfaces and strained films, where the low symmetry
crystal field splittings may not be negligible as compared to the exchange
field energies.

We acknowledge the NSRRC staff for providing us with an extremely stable beam.
We would like to thank Lucie Hamdan for her skillful technical and
organizational assistance in preparing the experiment. The research in Cologne
is supported by the Deutsche Forschungsgemeinschaft through SFB 608.


\begin{thebibliography}{99}
\bibitem{SOremark} Spin-orbit interaction in the $3d$ shell will not split the $^{3}A_{2}$
 ground state of a Ni $3d^{8}$ ion in $O_{H}$ symmetry. Spin-orbit coupling does split
 the $^{3}$B$_{1}$ state for a Ni ion in $D_{4h}$, but this splitting is negligibly small,
 of order $10^{-4}$ eV.
\end{thebibliography}
\end{document}